\documentclass[preprint,showpacs,preprintnumbers,amsmath,amssymb]{revtex4}

\usepackage{graphicx}
\usepackage{dcolumn}
\usepackage{bm}

\begin{document}

\title{Vortices in superconductors with a columnar defect:
finite size effects}

\author{Edson Sardella, Rodrigo de Alvarenga Freire, and P.\ N.\ Lisboa-Filho}
\affiliation{Departamento de F\'{\i}sica,
Faculdade de Ci\^encias, Universidade Estadual Paulista \\
Caixa Postal 473, 17033-360, Bauru-SP, Brazil}
\email{sardella@fc.unesp.br}

\date{\today}

\begin{abstract}
In the present work we investigate the behavior of a vortex in a long
superconducting cylinder near to a columnar defect at the center. The
derivations of the local magnetic field distribution and the
Gibbs free energy will be carried out for a cylinder and a 
cavity of arbitrary sizes. From the 
general expressions, it considered two particular limits: one in which
the radius of the cavity is very small but the radius of the
superconducting cylinder is kept finite; and
one in which the radius of the superconducting 
cylinder is taken very large (infinite)
but the radius of the cavity is kept finite. In both cases the
maximum number of vortices which are allowed in the cavity is determined. In addition, the surface barrier field for flux entrance into the cavity is calculated.
\end{abstract}
\maketitle
\section{Introduction}
The analysis of vortex motion in the presence of defects is one of the most important topic in the superconductivity research field. The relevance of this theme is based on the fact that the behavior of type-II superconductors in the presence of an applied magnetic field can be described by the motion and pinning of vortices in the material. 

In the last decades a number of magnetic anomalies, such as the Wohlleben effect, 
fish tail anomaly and the jumps on the magnetic response of mesoscopic samples, were reported relating the occurrence of defects and the granular conformance of both bulk samples and thin films \cite{passos,luzhbin}. Simultaneously, several initiatives using the most different theoretical approaches have been carried out to describe granular superconductors and their characteristic magnetic behavior \cite{fouad,deluca}. Today it is well known that all these features and anomalies are straight linked to the vortex dynamics.

Recently, the advances in the fabrication process related to nanotechnology made possible the production of mesoscopic rings \cite{zhang} and superconducting arrays with columnar defects \cite{moshchalkov}. Experiments using these nanoscopic samples showed that the connectivity and vorticity are quite different than those for the macroscopic ones. The physics revealed by those experiments has contributed to a major understanding of type-II granular superconductors. This has motivated
both experimental and theoretical physicists to investigate the
magnetic properties of either small superconductors or bulk samples with nanosized defects. The properties of the vortex lattice, for instance, in this superconductors of
confined geometries change radically with respect to the bulk superconductor ones.

One of the open questions within the framework of mesoscopic superconducting materials is the validity and complete understanding of the simple rule proposed by Mkrtchyan and Shmidt \cite{MS}. This rule indicates that the maximum number of vortices that enter a columnar defect is $n_s=r/2\xi$, where $r$ is the radius of the cylindrical cavity and $\xi$ is the superconducting coherence length. Although this rule has been questioned, it is a reasonable estimate of the cavity occupation \cite{doria1999,doria2000}.

In this paper the work of Refs.~\cite{MS,nordborg2000} is extended for the case of a long 
superconducting wire, of arbitrary size, with a columnar defect. The
applied magnetic field is parallel to 
the cylinder axis and the cylindrical cavity axis coincides with the
direction of the applied field. First we determine the local magnetic
field with appropriate boundary conditions and further, we calculate the
Gibbs free energy of an ensemble of vortices near the cavity with $n$
vortices inside it. On using the Gibbs free energy, the
force acting on a vortex near the cavity and near to the external surface of the superconductor is calculated. From the expression of the
force, we find a criterion for the maximum number of vortices allowed in
the cavity. It is shown that if the size of the cavity is not too
small, the saturation number differs substantially from the classical rule of
Mkrtchyan and Shmidt \cite{MS}. The surface barrier field for flux penetration into the cavity is also determined.

\section{The Magnetic Field}
The starting point of our study is the London equation for the
local magnetic field of a very strong type-II superconductor for
which the Ginzburg-Landau parameter $\kappa=\frac{\lambda}{\xi}\gg
1$. For our purposes, this equation can be more conveniently
written in cylindrical coordinates as
\begin{equation}\label{londonCilindricas}
-\lambda^2\left ( \frac{\partial^2
h}{\partial\rho^2}+\frac{1}{\rho}\frac{\partial
h}{\partial\rho}+\frac{1}{\rho^2}\frac{\partial^2
h}{\partial\varphi^2}\right
)+h=\Phi_0\,\sum_{i=1}^{N}\,\delta(\mbox{\boldmath
$\rho$}-\mbox{\boldmath $\rho$}_i)\;,
\end{equation}
where $\mbox{\boldmath $\rho$}_i$ is the position of the
$i$-vortex line, $\Phi_0$ is the quantum flux, and $N$ is the number of vortices outside the 
columnar defect.

Consider a long superconducting wire under an applied field $H$
parallel to the cylinder axis. We assume that there is a columnar
defect at the center of the cylinder. Let $a$ be the radius of the
cylinder and $r$ the radius of the cavity (columnar defect). This
is a Dirichlet problem in which the boundary conditions are given
by
\begin{equation}\label{Dirichlet}
h(a,\varphi)=H\;,\;\;\;h(r,\varphi)=H_0\;,
\end{equation}
where $H_0$ is the magnetic field inside the columnar defect. Due
to the symmetry of the geometry under investigation, the field in
the cavity is uniform and parallel to the cylinder axis. The
determination of $H_0$ will be discussed later on.

The solution of the London equation under these boundary
conditions may be found by means of the Green's function method.
We have for the local magnetic field
\begin{equation}\label{h}
h(\rho,\varphi)=\Phi_0\,\sum_{i=1}^{N}\,
G(\rho,\varphi,\rho_i,\varphi_i)+h_1(\rho)\;,
\end{equation}
where $G(\rho,\varphi,\rho^{\prime},\varphi^{\prime})$ is the
Green's function and $h_1(\mbox{\boldmath $\rho$})$ is the
solution of the homogeneous London equation as if no vortex were
present (Meissner state). By using Eqs.~(\ref{Dirichlet}), we can
easily determine the homogeneous solution,
\begin{equation}\label{h1}
h_1(\rho)=H_0\frac{
\Delta_0(\rho,a)}{\Delta_0(r,a)}+H\frac{\Delta_0(r,\rho)}{\Delta_0(r,a)}
\;.
\end{equation}
Here
\begin{equation}\label{Delta}
\Delta_m(x,y)=K_m(x/\lambda)I_m(y/\lambda)-K_m(y/\lambda)I_m(x/\lambda)\;,
\end{equation}
where $K_m(x)$ and $I_m(x)$ are the modified Bessel functions of
second kind.

The Green's function can be expanded in a Fourier series as
\begin{equation}\label{fgreen}
G(\rho,\varphi,\rho{\prime},\varphi{\prime})=\frac{1}{2\pi}\,
\sum_{m=-\infty}^{\infty}\,e^{im(\varphi-\varphi^{\prime})}
g_m(\rho,\rho^{\prime})\;.
\end{equation}

Although easy, it is very cumbersome to determine the Fourier
coefficients $g_m(\rho,\rho^{\prime})$. Here, we will omit this
calculation. The Dirichlet boundary conditions (\ref{Dirichlet})
impose that
\begin{equation}
g_m(r,\rho^{\prime})=g_m(a,\rho^{\prime})=0\;,
\end{equation}
at the surfaces, and
\begin{eqnarray}\label{boundarygm}
g_m((\rho^{\prime})^{-},\rho^{\prime}) & = &
g_m((\rho^{\prime})^{+},\rho^{\prime})\;,
\nonumber \\
\left ( \frac{\partial g_m}{\partial\rho} \right
)_{\rho=(\rho^{\prime})^{+}}- \left ( \frac{\partial
g_m}{\partial\rho}
 \right
 )_{\rho=(\rho^{\prime})^{-}} & = & -\frac{1}{\lambda^2\rho^{\prime}}\;.
\end{eqnarray}
The first condition represents the continuity of $g_m$ function at
$\rho=\rho^{\prime}$. The second condition represents the
discontinuity of the derivative of $g_m$ at $\rho=\rho^{\prime}$,
which comes from the delta function in
Eq.~(\ref{londonCilindricas}). By using these boundary conditions,
we find
\begin{equation}\label{gm}
g_m(\rho,\rho^{\prime})=\left \{
\begin{array}{ll}
\frac{\Delta_m(a,\rho^{\prime})\Delta_m(\rho,r)}
{\lambda^2\Delta_m(r,a)}\;,&\;\;\; r \le \rho \le \rho^{\prime} \\
\frac{\Delta_m(r,\rho^{\prime})\Delta_m(\rho,a)}
{\lambda^2\Delta_m(r,a)}\;, &\;\;\; \rho^{\prime} \le \rho \le
a\;,
\end{array}
\right .
\end{equation}
where $\Delta_m(x,y)$ is given by Eq.~(\ref{Delta}). Notice that
this function is antisymmetric by interchanging $x$ and $y$, that
is $\Delta_m(x,y)=-\Delta_m(y,x)$.

We still have to find the constant $H_0$. For that, we use the
supercurrent expression
\begin{equation}
\mbox{\boldmath $\nabla$}\times{\bf h}=\frac{1}{\lambda^2}\left(
\frac{\Phi_0}{2\pi}\mbox{\boldmath $\nabla$}\theta-{\bf
A}\right)\;,
\end{equation}
where ${\bf A}$ is the vector potential, and $\theta$ is the phase
of the order parameter. By integrating this equation in the circle
of radius $r$ containing the cavity and using the fact that the
phase is not single valued, we obtain
\begin{equation}\label{conditionH0}
-r\,\int_{0}^{2\pi}\,\left(\frac{\partial
h}{\partial\rho}\right)_{\rho=r}\,d\varphi=\frac{\Phi_0}{\lambda^2}n-\frac{\pi
r^2}{\lambda^2}H_0\;,
\end{equation}
where $n$ is the number of vortices which were captured by the
columnar defect. It should be clear the difference between $N$ and
$n$.

Only the $m=0$ term will survive in the integration of
Eq.~(\ref{conditionH0}). A length algebra leads us to the
following expression for the magnetic field in the cavity
\begin{equation}\label{H0}
H_0=\frac{\Phi_0}{2\pi\lambda r}\left[\frac{
n+\frac{1}{\Delta_0(r,a)}\,\sum_{i=1}^N\,\Delta_0(\rho_i,a)+\frac{2\pi\lambda^2H}{\Phi_0\Delta
_0(r,a)} } {
\frac{r}{2\lambda}+\frac{\delta_0(r,a)}{\Delta_0(r,a)} }
 \right]\;,
\end{equation}
where
\begin{equation}
\delta_m(x,y)=K_{m+1}(x/\lambda)I_m(y/\lambda)+K_m(y/\lambda)I_{m+1}(x/\lambda)\;.
\end{equation}

Notice that in the limit of large cylinder, $a\rightarrow\infty$
and $r$ finite, $\Delta_0(r,a)\rightarrow
K_0(r/\lambda)I_0(a/\lambda)$, and $\delta_0(r,a)\rightarrow
K_1(r/\lambda)I_0(a/\lambda)$. Then, we have
\begin{equation}
H_0=\frac{\Phi_0}{2\pi\lambda r}\left[\frac{
n+\frac{1}{K_0(r/\lambda)}\,\sum_{i=1}^N\,K_0(\rho_i/\lambda)} {
\frac{r}{2\lambda}+\frac{K_1(r/\lambda)}{K_0(r/\lambda)} }
 \right]\;,
\end{equation}
which is the result found recently in Ref.~\cite{nordborg2000}.

\section{The Gibbs Free Energy}
In order to proceed we need to evaluate the Gibbs free energy (per
unit length) which is given by
\begin{equation}\label{gbh}
{\sf G}={\sf F}-\frac{ABH}{4\pi}\;,
\end{equation}
where ${\sf F}$ is the London free energy (the Helmholtz free
energy in the thermodynamic context) and $B$ is the average
induction; here $A$ is the area of the cylinder cross section. In
the London approximation, the free energy is given by
\begin{equation}
{\sf F}=\frac{1}{8\pi}\,\int_0^{2\pi}\int_r^a\, \left [
\frac{\lambda^2}{\rho^2}\left ( \frac{\partial h}{\partial\varphi}
\right )^2+\lambda^2\left ( \frac{\partial h}{\partial\rho} \right
)^2 + h^2\right ] \,\rho\,d\rho\,d\varphi\;.
\end{equation}
By using integration by parts, we can write
\begin{equation}\label{Helmholtz}
{\sf F}  =
\frac{\Phi_0}{8\pi}\,\sum_{i=1}^{N}\,h(\rho_i,\varphi_i)+
\frac{\lambda^2aH}{8\pi}\,\int_0^{2\pi}\, \left (\frac{\partial
h}{\partial\rho}\right
)_{\rho=a}\,d\varphi-\frac{\lambda^2rH_0}{8\pi}\,\int_0^{2\pi}\,
\left (\frac{\partial h}{\partial\rho}\right
)_{\rho=r}\,d\varphi\;.
\end{equation}
Next, we compute the derivatives and insert into this last
equation. In addition, we use Eqs.~(\ref{h}) and (\ref{h1}). We
find
\begin{equation}\label{londonEnergy}
{\sf F}  =
\frac{\Phi_0^2}{8\pi}\,\sum_{i,j=1}^{N}\,G(\rho_i,\varphi_i,\rho_j,\varphi_j)+\frac{\lambda
aH^2}{4}\frac{\delta_0(a,r)}{\Delta_0(r,a)}-\frac{\lambda^2HH_0}{2}\frac{1}{\Delta_0(r,a)}
+\frac{\lambda rH_0^2}{4}\frac{\delta_0(r,a)}{\Delta_0(r,a)}\;.
\end{equation}

The average induction may be obtained by integrating the London
equation (\ref{londonCilindricas}) on the cross section. We have
\begin{eqnarray}\label{induction}
AB & = & N\Phi_0+\lambda^2\,\int_0^{2\pi}\,\left[ a\left (
\frac{\partial h}{\partial\rho}\right )_{\rho=a}-r\left
(\frac{\partial h}{\partial\rho}\right
)_{\rho=r}\right ]\,d\varphi \nonumber \\
& = & N\Phi_0+\frac{2\pi\lambda[aH\delta_0(a,r)+rH_0\delta_0(r,a))]}{\Delta_0(r,a)} \nonumber \\
& &
-\frac{2\pi\lambda^2(H+H_0)}{\Delta_0(r,a)}+\frac{\Phi_0}{\Delta_0(r,a)}\,\sum_{i=1}^{N}\,
[\Delta_0(\rho_i,r)-\Delta_0(\rho_i,a)]\;.
\end{eqnarray}

Finally, upon substituting Eqs.~(\ref{londonEnergy}) and
(\ref{induction}) into Eq.~(\ref{gbh}), we obtain for the Gibbs
free energy
\begin{eqnarray}\label{gibbsEnergy}
{\sf G} & = &
\frac{\Phi_0^2}{8\pi}\,\sum_{i,j=1}^{N}\,G(\rho_i,\varphi_i,\rho_j,\varphi_j)+\frac{\Phi_0H}{4
\pi\Delta_0(r,a)}\,\sum_{i=1}^{N}\,
[\Delta_0(\rho_i,a)+\Delta_0(r,\rho_i)]\nonumber \\ & &
+\frac{\lambda^2H^2}{2}\frac{1}{\Delta_0(r,a)}+\frac{\lambda
rH_0}{2}\frac{\delta_0(r,a)}{\Delta_0(r,a)}\left(\frac{H_0}{2}-H\right)-\frac{\lambda
aH^2}{4}\frac{\delta_0(a,r)}{\Delta_0(r,a)}-\frac{N\Phi_0H}{4\pi}\;.
\end{eqnarray}

In the Appendix we show that the Green's function can also be written as

\begin{eqnarray}\label{fgreenSim}
G(\rho,\varphi,\rho^{\prime},\varphi^{\prime}) & = &
\frac{1}{2\pi\lambda^2}K_0(|\mbox{\boldmath
$\rho$}-\mbox{\boldmath
$\rho$}^{\prime}|/\lambda) \nonumber \\
& & +\frac{1}{2\pi\lambda^2}\,\sum_{m=-\infty}^\infty\,\left [
K_m(a/\lambda)I_m(\rho_>/\lambda)\Delta_m(\rho_<,r) \right . \nonumber \\
& & \left . +K_m(\rho_>/\lambda)I_m(r/\lambda)\Delta_m(a,\rho_<)
\right ] \frac{\cos[m(\varphi-\varphi_j)]}{\Delta_m(r,a)}\;,
\end{eqnarray}
where $\rho_>$ and $\rho_<$ are defined in Eqs.~(\ref{rhomaior}) and
(\ref{rhomenor}). Notice that for $\mbox{\boldmath$\rho$}=\mbox{\boldmath$\rho$}^{\prime}$, London theory requires a regularization procedure. We use a sharp cutoff in which $|\mbox{\boldmath$\rho$}-\mbox{\boldmath$\rho$}^{\prime}|$ is replaced by $\xi$.

In some special cases in which we apply the above results, Eq.~(\ref{fgreenSim})
will be shown much more useful than Eqs.~(\ref{fgreen}) and (\ref{gm}). All the results above are general, so it should be applied to a
superconducting cylinder with a columnar defect, both with
arbitrary sizes. 

\section{Force}
We will be most interested in the particular case $N=1$. From
Eq.~\ref{gibbsEnergy}, we have
\begin{eqnarray}\label{gibbsEnergySingle}
{\sf G} & = &\frac{\Phi_0^2}{8\pi}G(\rho_1)+
\frac{\Phi_0H}{4
\pi\Delta_0(r,a)}
[\Delta_0(\rho_1,a)+\Delta_0(r,\rho_1)]\nonumber \\ & &
+\frac{\lambda rH_0(\rho_1)}{2}\frac{\delta_0(r,a)}{\Delta_0(r,a)}\left(\frac{H_0(\rho_1)}{2}-H\right) \nonumber \\ 
& & -\frac{\lambda
aH^2}{4}\frac{\delta_0(a,r)}{\Delta_0(r,a)}+\frac{\lambda^2H^2}{2}\frac{1}{\Delta_0(r,a)}-\frac{\Phi_0H}{4\pi}\;.
\end{eqnarray}

Now the Green's function is given by
\begin{eqnarray}\label{fgreenSingle}
G(\rho_1) & = &
\frac{1}{2\pi\lambda^2}K_0(\xi/\lambda) \nonumber \\
& & +\frac{1}{2\pi\lambda^2}\,\sum_{m=-\infty}^\infty\,\left [
K_m(a/\lambda)I_m(\rho_1/\lambda)\Delta_m(\rho_1,r) \right . \nonumber \\
& & \left . +K_m(\rho_1/\lambda)I_m(r/\lambda)\Delta_m(a,\rho_1)
\right ] \frac{1}{\Delta_m(r,a)}\;,
\end{eqnarray}
and the magnetic field in the cavity is
\begin{equation}\label{H0single}
H_0(\rho_1)=\frac{\Phi_0}{2\pi\lambda r}\left[\frac{
n+\frac{\Delta_0(\rho_1,a)}{\Delta_0(r,a)}+\frac{2\pi\lambda^2H}{\Phi_0\Delta
_0(r,a)} } {
\frac{r}{2\lambda}+\frac{\delta_0(r,a)}{\Delta_0(r,a)} }
 \right]\;.
\end{equation}

The force acting on a single vortex can be found by taking the
derivative of the Gibbs free energy. At the surface of the columnar
defect, the force (per unit length) is then given by
\begin{eqnarray}\label{forceCavity}
-f_r & = & \frac{\Phi_0^2}{8\pi}\left ( \frac{dG(\rho_1)}{d\rho_1}\right
)_{\rho_1=r}+\frac{\Phi_0H}{4
\pi\lambda\Delta_0(r,a)}
\left [ \frac{\lambda}{r}-\delta_0(r,a)\right ] \nonumber \\
& & +\frac{\lambda
rH_0(\rho_1)}{2}\frac{\delta_0(r,a)}{\Delta_0(r,a)}[H_0(r)-H]\left (
\frac{d H_0(\rho_1)}{d\rho_1}  \right )_{\rho_1=r}\;,
\end{eqnarray}
where the use of the identities $K_1(x)I_0(x)+K_0(x)I_1(x)=1/x$,
$K_0^{\prime}(x)=-K_1(x)$, and $I_0^{\prime}(x)=I_1(x)$ have been made.

Following a similar procedure, we can find an expression for the force
acting on a single vortex at the external surface of the superconductor. We have
\begin{eqnarray}\label{forceSuper}
-f_a & = & \frac{\Phi_0^2}{8\pi}\left ( \frac{dG(\rho_1)}{d\rho_1}\right
)_{\rho_1=a}+\frac{\Phi_0H}{4
\pi\lambda\Delta_0(r,a)}
\left [ \delta_0(a,r)-\frac{\lambda}{a} \right ] \nonumber \\
& & +\frac{\lambda
rH_0(\rho_1)}{2}\frac{\delta_0(r,a)}{\Delta_0(r,a)}[H_0(a)-H]\left (
\frac{d H_0(\rho_1)}{d\rho_1}  \right )_{\rho_1=a}\;.
\end{eqnarray}

The derivative of the Green's function can be easily evaluated on using
Eq.~(\ref{fgreenSingle}). Very close to the cavity, we can write
\begin{eqnarray}\label{fgreenSingleCavity}
G(\rho_1)  & = & \frac{1}{2\pi\lambda^2}\left[
K_0(\xi/\lambda)-\sum_{m=-\infty}^\infty\,K_m(\rho_1/\lambda)I_m(r/\lambda)\right
] \nonumber \\
& = & \frac{1}{2\pi\lambda^2}\left[
K_0(\xi/\lambda)-K_0((\rho_1-r)/\lambda)\right ]\;.
\end{eqnarray}
Notice that at $\rho_1=r+\xi$ the Green's function vanishes as required
by the boundary conditions. Consequently, on using a sharp cutoff we
have
\begin{equation}
\label{fgreenDerivativeCavity}
\left (
\frac{dG(\rho_1)}{d\rho_1}\right)_{\rho_1=r}=\frac{1}{\pi\lambda^2}\left
[ \frac{1}{2\xi} \right ]\;,
\end{equation}
where it has been used the approximation $K_1(\xi/\lambda)=\lambda/\xi$ which is valid for a strong type-II superconductor.

Similarly, we can find for the derivative of the Green's function at the
surface of the superconductor,
\begin{equation}
\label{fgreenDerivativeSuper}
\left (
\frac{dG(\rho_1)}{d\rho_1}\right)_{\rho_1=a}=-\frac{1}{\pi\lambda^2}\left
[ \frac{1}{2\xi} \right ]\;.
\end{equation}

Notice that these terms of both forces do not depend on the geometrical
factors, but only on the fundamental lengths.

From now on we will take two special cases. One case
will consist of a columnar defect very small compared to the penetration
length $\lambda$ but still keeping $a/\lambda$ finite. In the other case
we will take $r/\lambda$ finite and a very large cylinder,
$a/\lambda\rightarrow\infty$.

\subsection{Small Cavity, Finite Superconducting Cylinder}
We have experienced that to find a closed expression for the force, both
at the cavity and the external superconductor surface, is very cumbersome, even in
the case $r/\lambda\ll 1$. Nevertheless, by using symbolic mathematics
on computers this task becomes feasible \cite{program}. From
Eqs.~(\ref{fgreenDerivativeCavity}) and (\ref{forceCavityFinal}), up to second order in $r/\lambda$, we find
\begin{eqnarray}\label{forceCavityFinal}
-f_r & = & \frac{\Phi_0^2}{8\pi^2\lambda^2}\frac{1}{r}\left \{
\frac{r}{2\xi}-(n+1)+\left[ \left ( \frac{2\pi\lambda^2}{\Phi_0}H \right
) \left ( \frac{2}{I_0(a/\lambda)}-1 \right) \right. \right. \nonumber \\
& & \left. \left. - (n+1) \left (
C+\frac{K_0(a/\lambda)}{I_0(a/\lambda)}+\ln(r/2\lambda)\right )\right
]\frac{r^2}{2\lambda^2}\right \}\;,
\end{eqnarray}
where $C=0.577215\ldots$ is the Euler's constant.

Upon neglecting terms of second order, then the force will be negative
or vanish if $n+1\le r/2\xi$. Therefore, the maximum number of vortices
permitted into the cavity is given by
\begin{equation}
\label{saturationNumber}
n_s=n+1=\frac{r}{2\xi}\;.
\end{equation}

This result was firstly
found by Mkrtchyan and Shmidt \cite{MS}. However, their derivation was
valid only for infinite superconductors. It is remarkable that for
finite superconductors, this saturation number does not depend on the
radius of the superconducting cylinder but only of the size of the cavity. So, even in the case of a mesoscopic superconductor, in which $r<a\ll\lambda$, the proximity of
the external and internal surfaces has no influence on how the cavity
will be occupied.

Let us now turn our attention to the surface barrier field sufficient to
saturate the cavity with vortex entrance. From Eqs.~(\ref{forceSuper}) and
(\ref{fgreenDerivativeSuper}), by keeping only terms of second order or
less in $a/\lambda$, we find
\begin{equation}
\label{forcaSuperFinal}
-f_a=\frac{\Phi_0^2}{8\pi^2\lambda^2}\frac{1}{a}\left\{ \left ( \frac{2\pi\lambda^2}{\Phi_0}H \right
)\frac{a}{\lambda}\frac{I_1(a/\lambda)}{I_0(a/\lambda)}-n\frac{1}{I_0(a/\lambda)}-\frac{a}{2\xi} 
\right\}\;.
\end{equation}

If $-f_a>0$ the vortex enters the sample. Then, the minimum applied
magnetic field for entering $n+1$ vortices in the sample is
\begin{equation}
\label{surfaceField}
H_s=\left[ 1+\frac{2\xi}{a}\left( \frac{r}{2\xi}-1
\right)\frac{1}{I_0(a/\lambda)}
\right]\frac{I_0(a/\lambda)}{I_1(a/\lambda)}H_s^\infty\;,
\end{equation}
where $H_s^\infty=\Phi_0/4\pi\lambda\xi$ is the surface barrier field of a bulk
superconductor, and we have used Eq.~(\ref{saturationNumber}). Notice that
for $a/\lambda \gg 1$, $I_m(a/\lambda)=e^{a/\lambda}/\sqrt{2\pi
a/\lambda}$. Thus, for a large  superconductor, Eq.~(\ref{surfaceField})
reduces to $H_s=H_s^\infty$ which is a well known result for a bulk superconductor 
\cite{orlando}. On the other hand, if $a/\lambda\ll 1$, then
$I_0(a/\lambda)=1$, $I_1(a/\lambda)=a/2\lambda$, so that
\begin{equation}
\label{sufaceFieldMesoscopic}
H_s=\left[ 1+\frac{2\xi}{a}\left( \frac{r}{2\xi}-1
\right)
\right]\frac{2\lambda}{a}H_s^\infty\;.
\end{equation}

The term inside the square brackets is of order or larger then the unity,
so this field is much larger than the surface barrier field for a bulk
superconductor. As expected, a mesoscopic superconductor would required
a much larger applied field for flux penetration \cite{baelus2002}.

\subsection{Finite Cavity, Large Superconducting Cylinder}
Let us move on to the discussion of another special case in which the
size of the cavity is finite and the size of the superconducting cylinder is
infinite. Within these limits, the free energy of Eq.~(\ref{gibbsEnergySingle}) becomes
\begin{equation}\label{gibbsEnergySingleBulk}
{\sf G}=\frac{\Phi_0^2}{8\pi}G(\rho_1)+\frac{\lambda rK_1(r/\lambda)}{4K_0(r/\lambda)}[H_0(\rho_1)]^2-\frac{\Phi_0H}{4\pi}\;,
\end{equation}
where now
\begin{equation}
H_0(\rho_1)=\frac{\Phi_0}{2\pi\lambda r}
\left[
\frac{n+\frac{K_0(\rho_1/\lambda)}{K_0(r/\lambda)}}{
\frac{r}{2\lambda}+\frac{K_1(r/\lambda)}{K_0(r/\lambda)}}
\right]\;.
\end{equation}

Thus, the force acting on a single vortex at the surface of the cavity is given by
\begin{eqnarray}\label{forceCavityBulk}
-f_r & = & \frac{\Phi_0^2}{8\pi}\left ( \frac{dG(\rho_1)}{d\rho_1}\right
)_{\rho_1=r}+\frac{\lambda rK_1(r/\lambda)}{2K_0(r/\lambda)}H_0(r)\left (\frac{dH_0(\rho_1)}{d\rho_1}\right )_{\rho_1=r} \nonumber \\
& = & \frac{\Phi_0^2}{8\pi^2\lambda^2}\frac{1}{r}\left \{
\frac{r}{2\xi}-(n+1)\left [ 
\frac{\frac{K_1(r/\lambda)}{K_0(r/\lambda)}}{\frac{r}{2\lambda}+\frac{K_1(r/\lambda)}{K_0(r/\lambda)}}
\right ]^2
\right \}\;.
\end{eqnarray}

\begin{figure}[h]
\includegraphics[width=10cm]{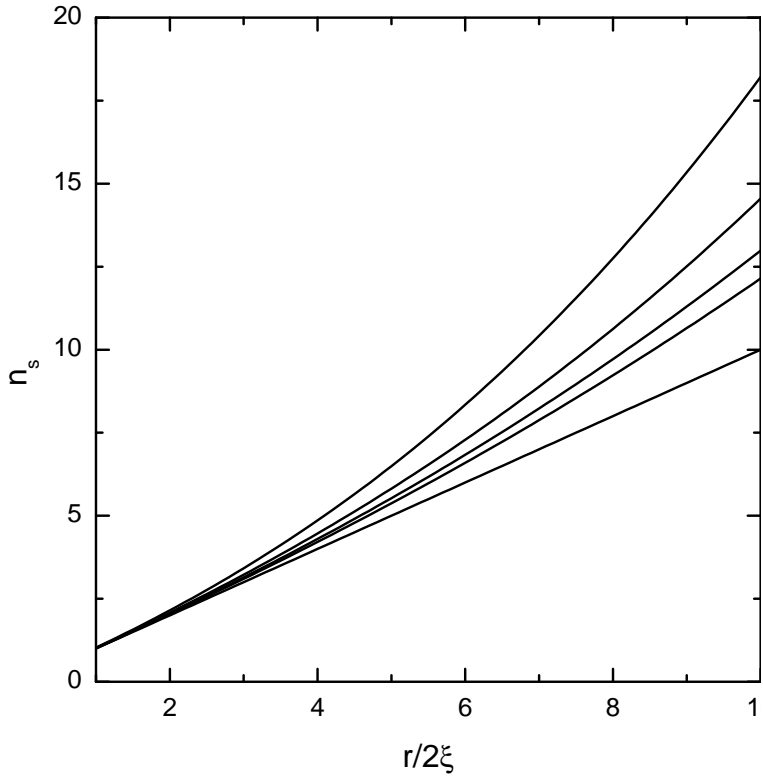}
\caption{The straight line is $n_s=r/2\xi$. The other curves are given
by Eq.~(\ref{saturationNumberBulk}); from the top to the bottom $\kappa=20,30,40,50$.}\label{fig1}
\end{figure}

The $(n+1)$-th vortex will be captured by the cavity if this number makes the force zero. Therefore, the saturation number is
\begin{equation}\label{saturationNumberBulk}
n_s=n+1=\frac{r}{2\xi}\left [ 1+\frac{r}{2\lambda}\frac{K_0(r/\lambda)}{K_1(r/\lambda)}\right ]^2\;.
\end{equation}
Notice that for the case of a small cavity this last expression yields
the Mkrtchyan-Shmidt rule. In Fig.~\ref{fig1} we plotted
Eq.~(\ref{saturationNumberBulk}) as a function of $r/2\xi$ for several
values of $\kappa$. We can see that, as the size of the cavity
increases, the saturation number differs significantly from the
Mkrtchyan-Shmidt classical rule. Thus, more vortices should be allowed
in the columnar defect if the radius of cavity increases, however the
larger $\kappa$ the lower the increasing is.

\section{Summary and Conclusions}
In summary, we have derived the local field and the free energy of an
ensemble of vortices around a columnar defect in a superconducting
wire for both the cavity and the superconductor of arbitrary sizes. We
also evaluated the force near to the cavity and external surfaces of the
superconductor. It has also been found that for a large superconductor, the cavity saturates at a
larger number of vortices as its size increases. However, not linearly in
$r/2\xi$ as predicted in \cite{MS}. In fact, in
\cite{doria1999,doria2000} it was presented some numerical simulations
which states that the population of the columnar defect grows not
linearly with $r/2\xi$, although in a different manner as found in the present work. Perhaps because in the present work it has been used the London theory which is strongly dependent on the cutoff model for the vortex core, whereas in those references the Ginzburg-Landau theory was employed.

\appendix

\section{Green's Function}
The expression found for the Fourier coefficients (\ref{gm}) may
be written in a more compact form as
\begin{equation}
g_m(\rho,\rho^{\prime})=
\frac{\Delta_m(a,\rho_>)\Delta_m(\rho_<,r)}{\lambda^2\Delta_m(r,a)}\;.
\end{equation}
where
\begin{eqnarray}\label{rhomaior}
\rho_>= \left\{
\begin{array}{ll}
\rho^{\prime}&\;,\;\;\;\rho<\rho^{\prime} \\
\rho         &\;,\;\;\;\rho>\rho^{\prime}\;,
\end{array}
\right.
\end{eqnarray}
and
\begin{eqnarray}\label{rhomenor}
\rho_<= \left\{
\begin{array}{ll}
\rho^{\prime}&\;,\;\;\;\rho>\rho^{\prime} \\
\rho         &\;,\;\;\;\rho<\rho^{\prime}\;,
\end{array}
\right.
\end{eqnarray}

Introducing the expressions of $\Delta_m$ of Eq.~(\ref{Delta}), we
find
\begin{eqnarray}
g_m(\rho,\rho^{\prime}) & = & \frac{1}{\lambda^2\Delta_m(r,a)}
\left [
K_m(a/\lambda)I_m(r/\lambda)I_m(\rho_>/\lambda)K_m(\rho_</\lambda)\right
.
\nonumber \\
& & \left .
-K_m(r/\lambda)K_m(a/\lambda)I_m(\rho_>/\lambda)I_m(\rho_</\lambda)\right
. \nonumber \\
& & \left .
-I_m(r/\lambda)I_m(a/\lambda)K_m(\rho_>/\lambda)K_m(\rho_</\lambda)\right
.
\nonumber \\
& &\left . +
K_m(r/\lambda)I_m(a/\lambda)K_m(\rho_>/\lambda)I_m(\rho_</\lambda)
\right ]\;.
\end{eqnarray}

Inside the square brackets we can sum and subtract the following
term
\[
K_m(a/\lambda)I_m(r/\lambda)K_m(\rho_>/\lambda)I_m(\rho_</\lambda)\;,
\]
and obtain
\begin{eqnarray}
g_m(\rho,\rho^{\prime}) & = &
\frac{1}{\lambda^2}K_m(\rho_>/\lambda)I_m(\rho_</\lambda)
\nonumber \\
& & +\frac{1}{\lambda^2\Delta_m(r,a)}\left [
K_m(a/\lambda)I_m(r/\lambda)I_m(\rho_>/\lambda)K_m(\rho_</\lambda)\right
. \nonumber \\
& & \left . -
K_m(a/\lambda)K_m(r/\lambda)I_m(\rho_>/\lambda)I_m(\rho_</\lambda)
\right . \nonumber \\
& & \left . -
I_m(a/\lambda)I_m(r/\lambda)K_m(\rho_>/\lambda)K_m(\rho_</\lambda)\right
. \nonumber \\
& & \left . +
K_m(a/\lambda)I_m(r/\lambda)K_m(\rho_>/\lambda)I_m(\rho_</\lambda)\right
]\;.
\end{eqnarray}

Substituting this equation in Eq.~(\ref{fgreen}) and using the
identity
\begin{equation}
\sum_{m=-\infty}^\infty\,K_m(x)I_m(y)\cos(m\varphi)=
K_0(\sqrt{x^2+y^2-2xy\cos\varphi}\,)\;,
\end{equation}
we obtain
\begin{eqnarray}
G(\rho,\varphi,\rho^{\prime},\varphi^{\prime}) & = &
\frac{1}{2\pi\lambda^2}K_0(|\mbox{\boldmath
$\rho$}-\mbox{\boldmath $\rho$}^{\prime}|/\lambda)+
\frac{1}{2\pi\lambda^2}\,\sum_{m=-\infty}^\infty\,\left \{ 
K_m(a/\lambda)I_m(\rho_>/\lambda)\right. \nonumber \\
& & \left. \times \left [K_m(\rho_</\lambda)I_m(r/\lambda)-
K_m(r/\lambda)I_m(\rho_</\lambda)\right ]\right . \nonumber \\
& & \left . +K_m(\rho_>/\lambda)I_m(r/\lambda)\right. \nonumber \\
& & \left. \times\left [ K_m(a/\lambda)I_m(\rho_</\lambda)-
K_m(\rho_</\lambda)I_m(a/\lambda)\right ] \right \} \nonumber \\
& & \times\frac{\cos[m(\varphi-\varphi^{\prime})]}{\Delta_m(r,a)}\;.
\end{eqnarray}

On using the definition (\ref{Delta}) we arrive at Eq.~(\ref{fgreenSim}).


\begin{thebibliography}{}
\bibitem{program} The authors developed a program by using symbolic mathematic on computers which calculates the expansion of the force in a power series of $r/\lambda$. For those interested, this program is available directly from the authors.
\end{thebibliography}
\end{document}